\documentclass[trackchanges,twocolumn]{aastex7}

\usepackage{url}

\begin{document}

\title{Spiral arms across cosmic time: JWST measurements of the pitch angles of spiral galaxies at $z<3.5$}

\author[orcid=0009-0002-7605-3478,sname='North America']{Vicki Kuhn}
\affiliation{Department of Physics and Astronomy, University of Missouri, Columbia, MO 65211, USA}
\email[show]{vkuhn@missouri.edu}

\author[orcid=0000-0003-2775-2002,sname='North America']{Yicheng Guo}
\affiliation{Department of Physics and Astronomy, University of Missouri, Columbia, MO 65211, USA}
\email[show]{guoyic@missouri.edu} 

\author{Sophie Rentschler}
\affiliation{Department of Physics and Astronomy, University of Missouri, Columbia, MO 65211, USA}
\email{sgrnkg@missouri.edu}

\author{Maxmillian Castillo}
\affiliation{Department of Physics and Astronomy, University of Missouri, Columbia, MO 65211, USA}
\email{mackpc@missouri.edu}

\author[orcid=0009-0002-0543-8880]{Gourab Nandi}
\affiliation{Department of Physics and Astronomy, University of Missouri, Columbia, MO 65211, USA}
\email{g.nandi@missouri.edu}

\author{Ellie Dugdale}
\affiliation{Department of Physics and Astronomy, University of Missouri, Columbia, MO 65211, USA}
\email{erdhh7@missouri.edu}

\author{Tsinat Mitiku}
\affiliation{Department of Physics and Astronomy, University of Missouri, Columbia, MO 65211, USA}
\email{tmmvft@missouri.edu}

\begin{abstract}

The properties of spiral galaxies in the early universe remain poorly studied and, as such, little is known about their nature and evolution. We use JWST data to measure the pitch angles of spiral galaxies across cosmic time. Our sample consists of 593 spiral galaxies with stellar masses ($M_*$) greater than $10^{10} M_\odot$ up to $z \sim 3.5$, drawn from the CEERS and JADES surveys. Spiral galaxies are identified by fine-tuning a Zoobot deep-learning model. We use SpArcFiRe to identify spiral arms and measure their pitch angles. We find no significant redshift evolution in the average pitch angle across the full sample. However, in the most massive systems (log$(M_*/M_\odot)=11-12$), spiral arms slightly wind up with time. We show that at $z>1.25$, pitch angle does not correlate with some key internal galaxy properties (stellar mass, bulge mass, disk mass, specific star formation rate [sSFR]). In contrast, at $z<1.25$, pitch angle shows a weak but statistically significant negative correlation with stellar mass, bulge mass, and disk mass, and a positive correlation with sSFR at $z<0.75$. We also find no dependence of pitch angle on the tidal strength applied by nearby companions. These results indicate a transition epoch at $z\sim1$: above this redshift, spiral structures appear to be primarily locally driven and not correlated with global galaxy properties; and below this redshift, spiral arms are regulated by global gravitational potential, consistent with the predictions of the density wave theory.
\end{abstract}



\section{Introduction} 
Spiral structure is one of the most prominent and eye-catching out of any galactic features. Galaxies with spiral structures make up the majority of high-mass galaxies at low-redshift \citep{lintott2011,willett2013}. The arms are sites of star formation and dust (e.g., \cite{calzetti2005,holwerda2005}). Spiral arms are typically classified into three main categories: grand-design, multi-armed, and flocculent \citep{elmegreen1982}. Grand-designs are characterized by long, symmetric arms, while flocculents have short, patchy, and fragmented arms, which are more common in faint galaxies \citep{elmegreen1985}. Multi-armed galaxies are an in-between of the two, having both long and short arms that are non-symmetric. In the local universe, grand-design and multi-armed spiral galaxies make up the majority (e.g., \cite{diaz_garcia2019,wei2024}).

The formation of spiral arms has been a major debate for many decades. There are several theories that try to explain the existence and maintenance of spiral arms with no clear consensus in the literature. The most notable one is the density wave theory \citep{lin_shu1964}. It proposes that spiral structure consists of long-lived quasi-stationary standing waves that rotate through the disk at a constant pattern speed. In this framework, stars and gas pass through these regions, which experiences an increase in density. On the other hand, spiral arms could be produced through swing amplification \citep{goldreich1965,julian1966}, where a leading perturbation in a rotating disk is sheared into a trailing pattern, amplifying the pattern. Arms produced this way can be short-lived, transient, and recurrent. This is supported by N-body simulations (e.g., \cite{sellwood1984,fujii2011,donghia2013}), which do not appear to be able to produce long-lived arms \citep{dobbs2014}. Swing amplified arms typically appear in isolated environments and are multi-armed (e.g., \cite{donghia2013}).

Other theories include the manifold theory, where spiral arms are composed of groups of stars which originate at the ends of bars \citep{athanassoula2009a}, though the bar has been found to not be the main driver of spiral structure (e.g., \cite{bittner2017,diaz_garcia2019}). Spiral arms can also be induced through interactions \citep{toomre1972}.

The tightness of the spiral arms has historically been a major factor in determining Hubble type \citep{hubble1926,sandage1961}, along with the prominence of the bulge. This tightness is measured by the pitch angle, the angle between two tangent lines to a spiral arm and circle around the galactic center. Tightly wound arms correspond to small pitch angle values and loosely wound arms correspond to large pitch angle values.

Correlations between the pitch angle and various galaxy properties can provide insight into the formation mechanisms of spiral arms. There have been many studies conducted on low-redshift spirals with varying results. \cite{yu2019} found that pitch angle decreases with mass (total stellar, bulge, and disk). The negative correlation between pitch angle and concentration further supports the density wave theory \citep{lin_shu1964,roberts1975,savchenko2013,yu2019}. Density wave theory purports that the thickness of the disk and size of the bulge should have an effect on the tightness of the arms \citep{lin_shu1966} which has been seen observationally at $z\sim0$ (e.g., \cite{savchenko2013}).

On the other hand, \cite{diaz_garcia2019} found no dependence of the pitch angle on bulge mass while \cite{lingard2021} observed no correlation with bulge size. Furthermore, \cite{lingard2021} found large variations in the pitch angle which lends support to the scenario that arms wind up. While \cite{hart2017} reported a weak negative correlation between pitch angle and stellar mass, \cite{yu2019} found a stronger negative correlation, which they attribute to their differences in pitch angle measurements or sample selection. Our work will explore spiral galaxies at higher redshifts to test if there are any correlations between pitch angle and galaxy properties early on in spiral formation.

Before 2021, studies on the structure and other parameters of spiral galaxies at high-redshift were infrequent. Only a handful of spiral galaxies were found and spectroscopically confirmed at $z>2$ with the highest being $z\sim4.41$ \citep{law2012,yuan2017,tsukui2021,wu2023}. \cite{elmegreen2014} were even able to find all arm types at $z\sim1$. JWST has cleared a path to study galaxies at high-redshift. Within the first two years, multiple studies found a surprising discovery: disk galaxies dominate at high-redshift (e.g., \cite{ferreira2023,kartaltepe2023,huertas_company2024}). Furthermore, several studies have found bars and spiral features out to $z\sim3$ (e.g., \cite{costantin2023,guo2023,leconte2024,xiao2025}). However, few studies have looked at high-redshift spirals, and only recently has it become evident that more than a handful of spirals existed in the early universe (e.g., \cite{kuhn2024,espejo_salcedo2025}). 

Only a couple of studies have measured the pitch angles of high-redshift spiral galaxies. Recent studies \citep{reshetnikov2022,reshetnikov2023,chugunov2025} have shown an evolution of the pitch angle, with values becoming larger with redshift. This suggests that spiral arms wind up and become tighter over time, consistent with being transient in nature and in contrast to the density wave theory, which predicts a constant pitch angle. This is further justified by their results from the Pringle-Dobbs test \citep{pringle2019}; they find that the distribution of the cotangent of the pitch angle is fairly uniform indicating that the pitch angle changes over time. While these results are interesting if true, the sample sizes are small, especially at $z>2$. The distribution of the cotangent of the pitch angle would benefit with a larger sample.

In this paper, we measured the pitch angles of several hundred high-redshift galaxies and investigated how this is correlated with various properties of spiral galaxies, such as stellar mass and specific star formation rate (sSFR). We used observations from JWST and only focused on the highest-mass galaxies. We also categorized the spiral arms and briefly looked at their environments. We aim to understand what factors play the more prominent roles in the formation of spiral structure and at what time we see changes taking place.

The structure of this paper is as follows. Section \ref{sec:data} describes the sample of data and pitch angle measurements. In Section \ref{sec:results}, we present our results of the pitch angles as functions of various properties and in Section \ref{sec:discussion} we provide a discussion of our results. We summarize our results in Section \ref{sec:summary}. We adopt a $
\Lambda$CDM cosmology with $H_0=70$, $\Omega_m=0.3$, and $\Omega_\Lambda=0.7$.


\section{Sample} \label{sec:data}
\subsection{Imaging data and galaxy catalog}
We used data from the publicly released CEERS \citep{bagley2023,finkelstein2025,10.17909/z7p0-8481} and JADES \citep{eisensteina2023,eisensteinb2023,rieke2023,10.17909/8tdj-8n28} surveys. CEERS has ten pointings in 7 wavebands and covers $\sim$100 arcmin$^2$ of the sky. Mosaics were downloaded from their website\footnote{\url{https://ceers.github.io/index.html}} and pointings 1, 2, 3, and 6 were released in v0.5 and pointings 4, 5, 7, 8, 9, and 10 in v0.6. JADES has one mosaic each for GOODS-S and GOODS-N in 14 filters and a total coverage of $\sim$175 square arcmin$^2$. We obtained the mosaics from their website\footnote{\url{https://archive.stsci.edu/hlsp/jades}} where we used v1.0 for GOODS-N and v2.0 for GOODS-S.

\begin{figure*}[ht!]
\plotone{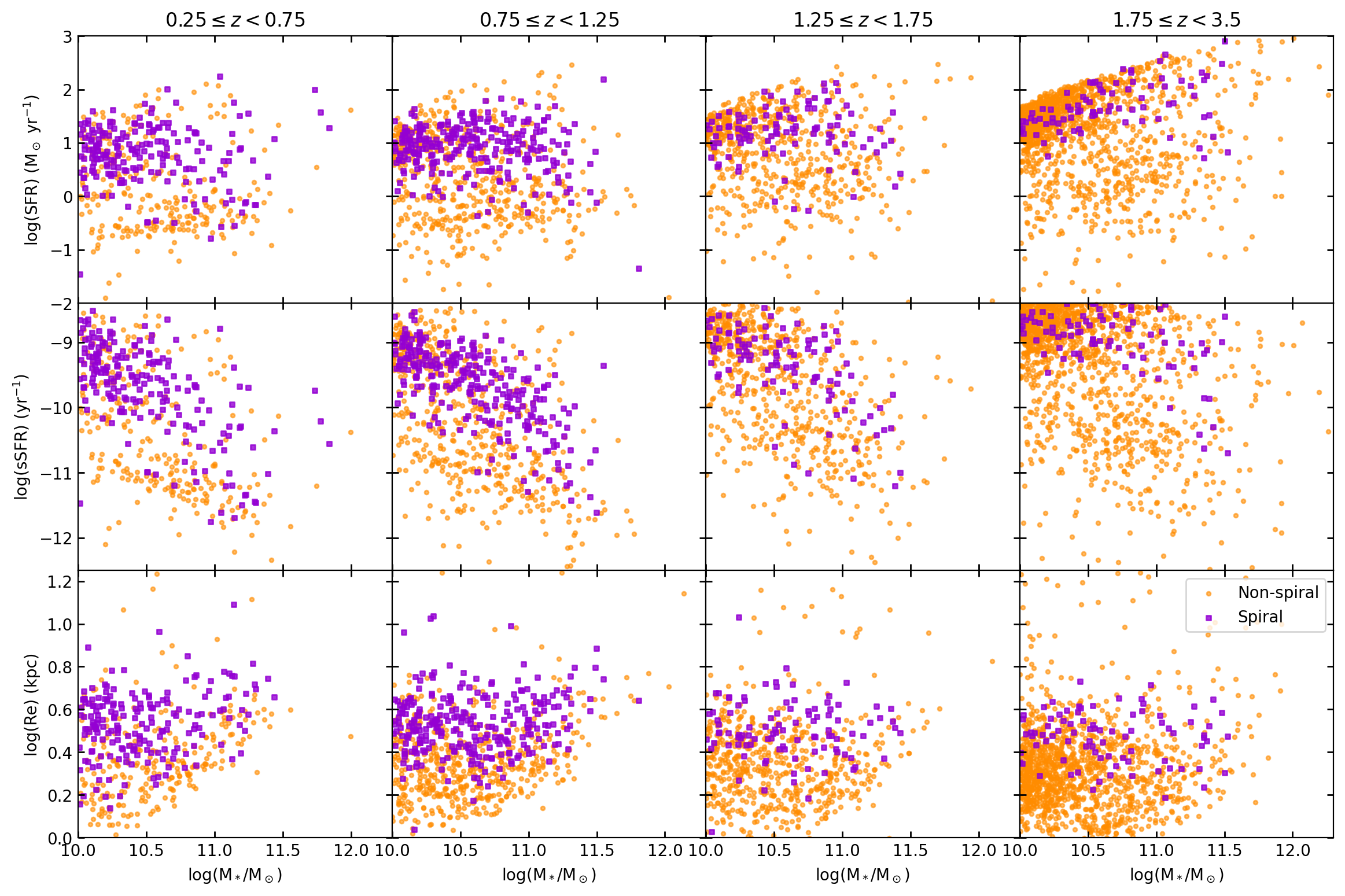}
\caption{Top row: Our sample in the SFR--stellar mass diagram at different redshift bins. Spiral galaxies are shown in purple squares and non-spiral galaxies are shown in orange circles. Middle row: Similar to the top panel, sSFR--stellar mass diagram. Bottom row: The half-light radius--stellar mass diagram. Spiral galaxies tend to have higher SFRs, sSFRs, and sizes compared to non-spiral galaxies.
\label{fig:fig1}}
\end{figure*}

We constructed our parent sample from the \textsc{ASTRODEEP} photometric catalog \citep[e.g.,][]{merlin2024}, which combined deep \textit{HST} and \textit{JWST} imaging across some major extragalactic legacy fields. The catalog provided homogeneous multi-wavelength photometry spanning approximately 0.4--4.4$\mu$m, enabling robust measurements of the spectral energy distributions (SEDs) for galaxies across a broad redshift range. In the catalog, sources were detected on stacks of NIRCam 3.56 and 4.44 $\mu$m images, and photometry was measured using PSF-matched images to unify angular resolutions among the imaging datasets.

The \textsc{ASTRODEEP} catalog also compiled an extensive collection of spectroscopic redshifts gathered over more than two decades from the literature, combining measurements from approximately 50 spectroscopic studies across these survey fields. For galaxies lacking secure spectroscopic redshifts, photometric redshifts were measured by \textsc{ASTRODEEP}. We chose the values measured using the \textsc{zphot} code \citep{fontana2000}. The photometric redshifts were calibrated against the available spectroscopic samples and achieve accuracies comparable to those obtained in other deep extragalactic surveys, with a scatter (standard deviation) of the relative error $\Delta z/(1+z) \sim 0.04$ and a low  outlier fraction ($\sim6\%)$. Throughout this paper, we adopt the spectroscopic redshifts whenever available and otherwise use the photometric redshifts.

Stellar mass ($M_*$) was derived by fitting the observed multi-wavelength photometry at the adopted redshift using \textsc{EAzY} \citep{Brammer2008}. We used the standard set of 12 ``tweaked'' FSPS templates distributed with the code, which are based on the Flexible Stellar Population Synthesis \citep[FSPS;][]{Conroy2009,ConroyGunn2010} models but empirically adjusted to better reproduce the observed colors of galaxies over a wide range of redshifts and star-formation histories. The fitting procedure determines the template combination that best reproduces the observed SED, from which $M_*$ and rest-frame colors are inferred. 

Star-formation rates (SFRs) were estimated using the rest-frame UVJ colors derived from \textsc{EAzY}. Following \citet{fang2018}, we projected galaxies onto a rotated coordinate system in the UVJ plane whose axes are defined parallel and perpendicular to the star-forming sequence. The perpendicular coordinate correlates strongly with specific star-formation rate (sSFR) and provides an empirical sSFR estimate with a typical uncertainty of $\sim$0.2 dex.

Galaxy size was obtained from the \textsc{ASTRODEEP} catalogs using the \texttt{FLUX\_RADIUS} parameter (the half-light radius) measured by \textsc{SExtractor} \citep{bertin1996} on the combined F356W+F444W detection images. The reported radii were converted from pixel units to physical units (kpc) using the adopted galaxy redshifts and the corresponding angular-diameter distances.

The bulge-to-total light ratio ($B/T$) was taken from the DAWN JWST Archive \citep{genin2025}, which performed bulge--disk image decomposition in the F444W band. We used $B/T$ to estimate bulge and disk stellar masses. Specifically, the stellar mass of the bulge ($M_b$) was calculated as $M_b = (B/T) \times M_*$, where $M_*$ is the total stellar mass derived from the SED fitting described above, and the stellar mass of the disk ($M_d$) was then calculated as $M_d = M_* - M_b$.


\subsection{Sample selection}

From the parent catalog, we selected massive galaxies with $M_*\geq10^{10} (M_\odot)$ in the redshift range of 0 $\leq$ z $<$ 5, resulting in 4271 galaxies.

We used Zoobot \citep{walmsley2023}, a python package that uses a convolutional neural network trained to predict galaxy morphology, to identify spiral galaxies. We use the pre-trained model ConvNeXT-Base model \citep{liu2022}, which uses 88.6M parameters. We fine-tuned this model with a sample of visually identified spiral galaxies in CEERS based on their rest frame V-band images (see \cite{kuhn2024}). In that sample, each galaxy had a score ranging from 0 to 6, reflecting how many people saw spiral structure (0 - no one saw spiral structure and 6 - everyone saw spiral structure). We put galaxies with $\geq4$ votes in the spiral category and $\leq3$ votes in the non-spiral category. Zoobot outputs a probability distribution on a scale from 0 to 1 which predicts the model's confidence in its prediction. We categorized galaxies as spirals with the distribution value $>0.5$ which resulted in 657 spiral galaxies. Figure \ref{fig:fig1} shows the SFR, sSFR, and half-light radius of our spiral and non-spiral sample. About 2/3 of galaxies in our spiral sample have spectroscopic redshifts.

\subsection{Pitch angle measurements}
There are several methods of measuring the pitch angle. These methods include one-dimensional Fourier analysis (e.g., \cite{kendall2011,yu2018}), two-dimensional Fourier analysis (e.g., \cite{davis2012}), photometric decomposition \citep{chugunov2025}, manual tracing (e.g., \cite{diaz_garcia2019}), \textsc{Spirality} \citep{shields2022}, SpArcFiRe \citep{davis2014}, and several others.


We used SpArcFiRe to measure our pitch angles. SpArcFiRe is an entirely automated program that, given an image, will center and de-project the galaxy, then extracts information about the spiral arms. Rather than measuring entire spiral arms, SpArcFiRe fits logarithmic spiral arcs to each arm segment found based on orientations.

\begin{figure*}[ht]
\includegraphics[width=1\linewidth]{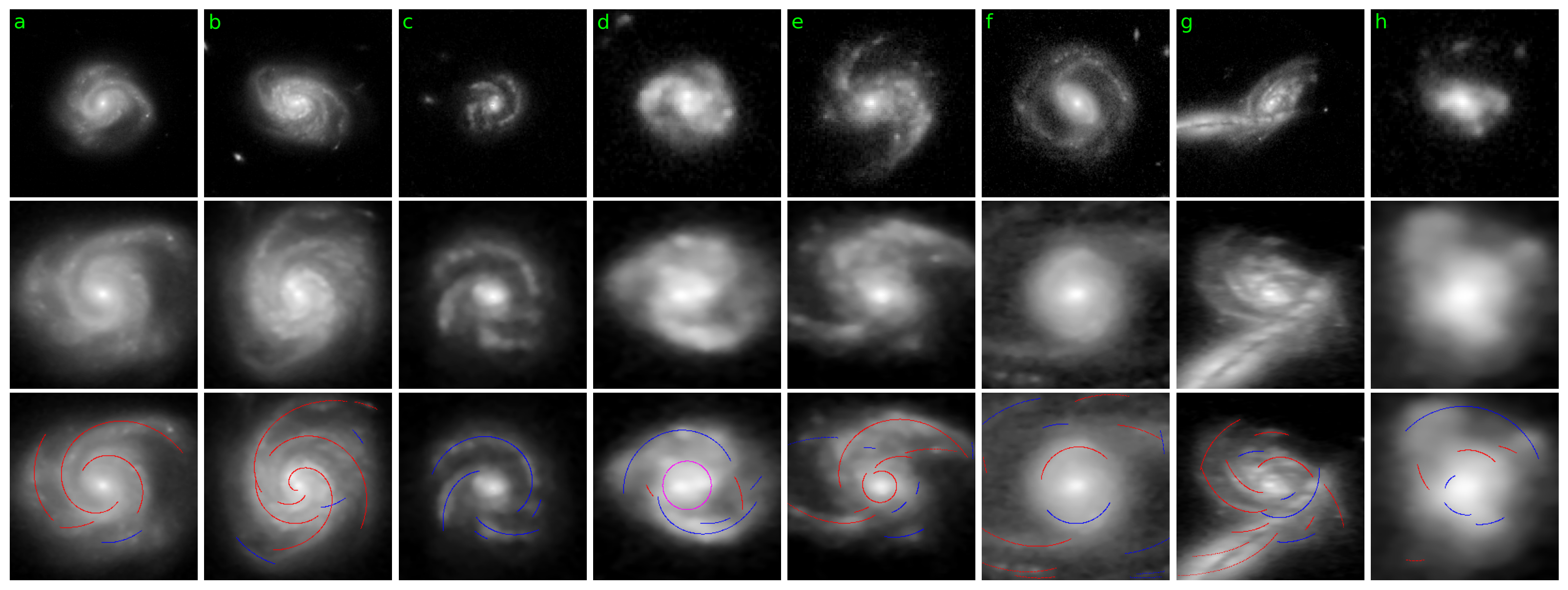}
\caption{Example SpArcFiRe outputs. Top row: The original image that was presented. Middle row: The centered and de-projected image. Bottom row: The middle image with final arcs superimposed. (a)-(d) Spiral galaxies at $z=0.284, 1.098, 2.033, 3.225$, respectively. (e) A bulge being improperly fit as an arc. (f) A spiral galaxy that has been zoomed in too close. (g) A pair of galaxies merging. (h) A non-spiral being fitted with arcs.
\label{fig:sparcfire}}
\end{figure*}

Once the galaxies were run through SpArcFiRe, they were visually inspected to check for fitting issues and various other issues. We identified galaxies that were not visually identified as spirals (5), galaxies that were in the process of merging (13), and arms that were cut off during the standardization step (36). For the last one, we resized the images and ran SpArcFiRe on them again (18). Though SpArcFiRe is automated, there are parameters the user can change before a fitting is done. These include parameters such as radially blurring, increase in contrast, minimum brightness for an arm segment to be an arc, among others. To see how much the fitting changed, we ran SpArcFiRe with two settings: the first with the default parameters and the second with changed "mergeChkMinClusSz" and "unsharpMaskAmt" values. The second version returned more galaxies with smaller arcs and overall a poor job of finding arcs. A common issue was SpArcFiRe identifying the bulge as an arc or mis-identifying non-arms as arcs (see Fig. \ref{fig:sparcfire}). These were flagged and are included in the final catalog. For galaxies where SpArcFiRe was unable to find arms, these are still included in the catalog though excluded from analysis (28). Overall, our sample includes 593 spiral galaxies with pitch angle measurements.

While \cite{davis2014} suggests using a threshold length of 75 pixels, we include every arc above 50 pixels. We found both the average pitch angle (only with arms $>75$ pixels in length) and the length weighted pitch angle mean for each galaxy (we used this value in our analysis). Uncertainties in the pitch angle were derived as the length-weighted standard deviation measured from individual arcs. Therefore, arcs that are longer have a greater weight. Typical uncertainties are $\sim8^\circ$.


\subsection{Pitch angle comparisons}
\begin{figure}[h]
\includegraphics[width=1\linewidth]{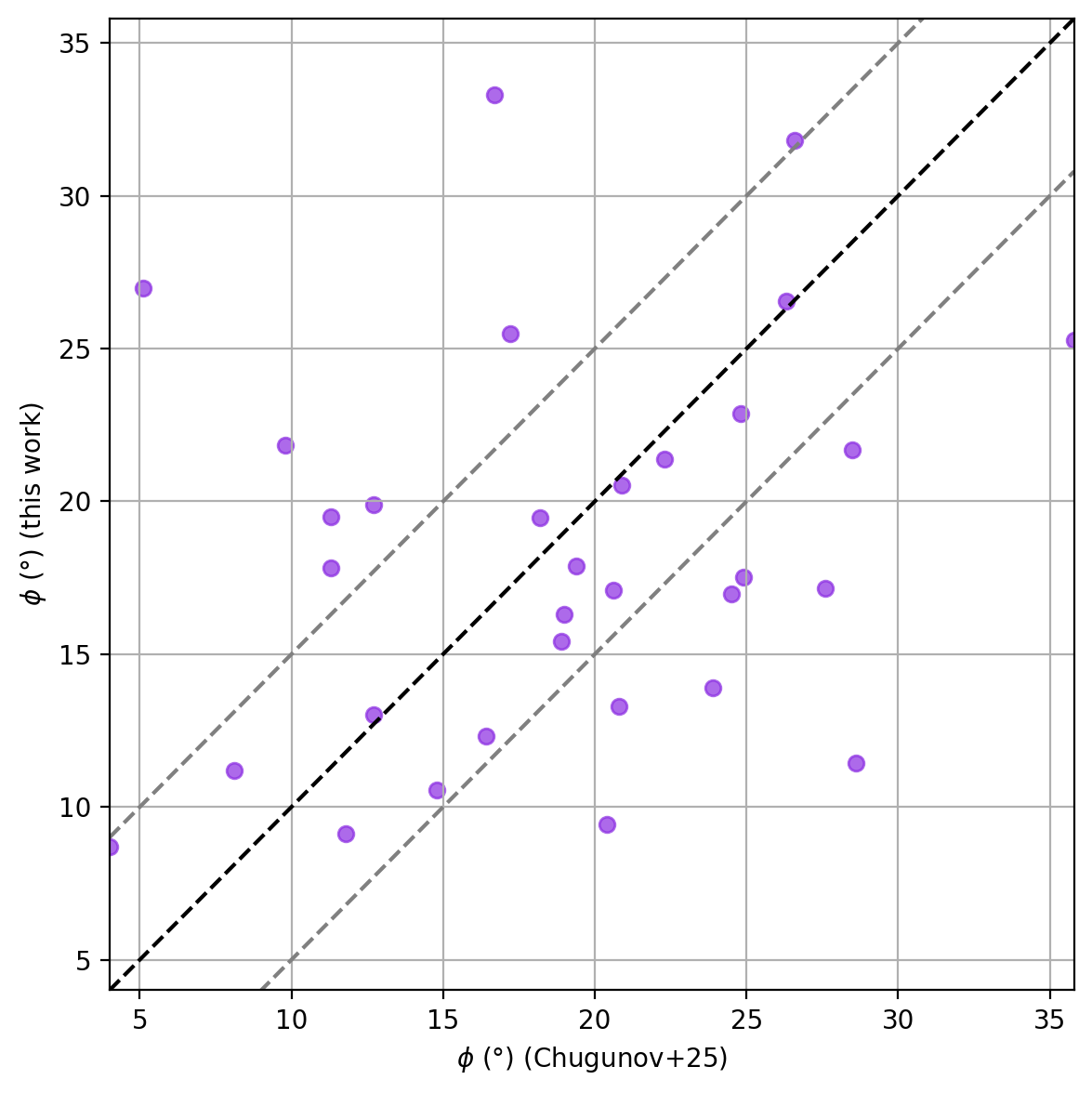}
\caption{A comparison between our SpArcFiRe measurements and those of \cite{chugunov2025} who used a photometric decomposition method. The mean offset is $-0.57^\circ$ and scatter of $\sigma=8.31^\circ$ between our two measurements.
\label{fig:comparison}}
\end{figure}

In this section we compare our pitch angle measurements with those from the literature, namely \cite{chugunov2025}. \cite{chugunov2025} had 33 galaxies from CEERS and JADES and we obtained the images they used in their analysis from their website (using the filter closest to the rest-frame V band based on their redshift)\footnote{\url{https://github.com/IVChugunov/Distant_spirals_decomposition}}. We show a comparison between their pitch angle values using their photometric decomposition method and our method of SpArcFiRe (Figure \ref{fig:comparison}). We were able to obtain pitch angles from 32 galaxies (purple circles). The black dashed line shows the 1:1 ratio and the gray lines show $\pm5^\circ$.


\section{results} \label{sec:results}
\subsection{Pitch angle vs. redshift}

\begin{figure*}[ht]
\includegraphics[width=1\linewidth]{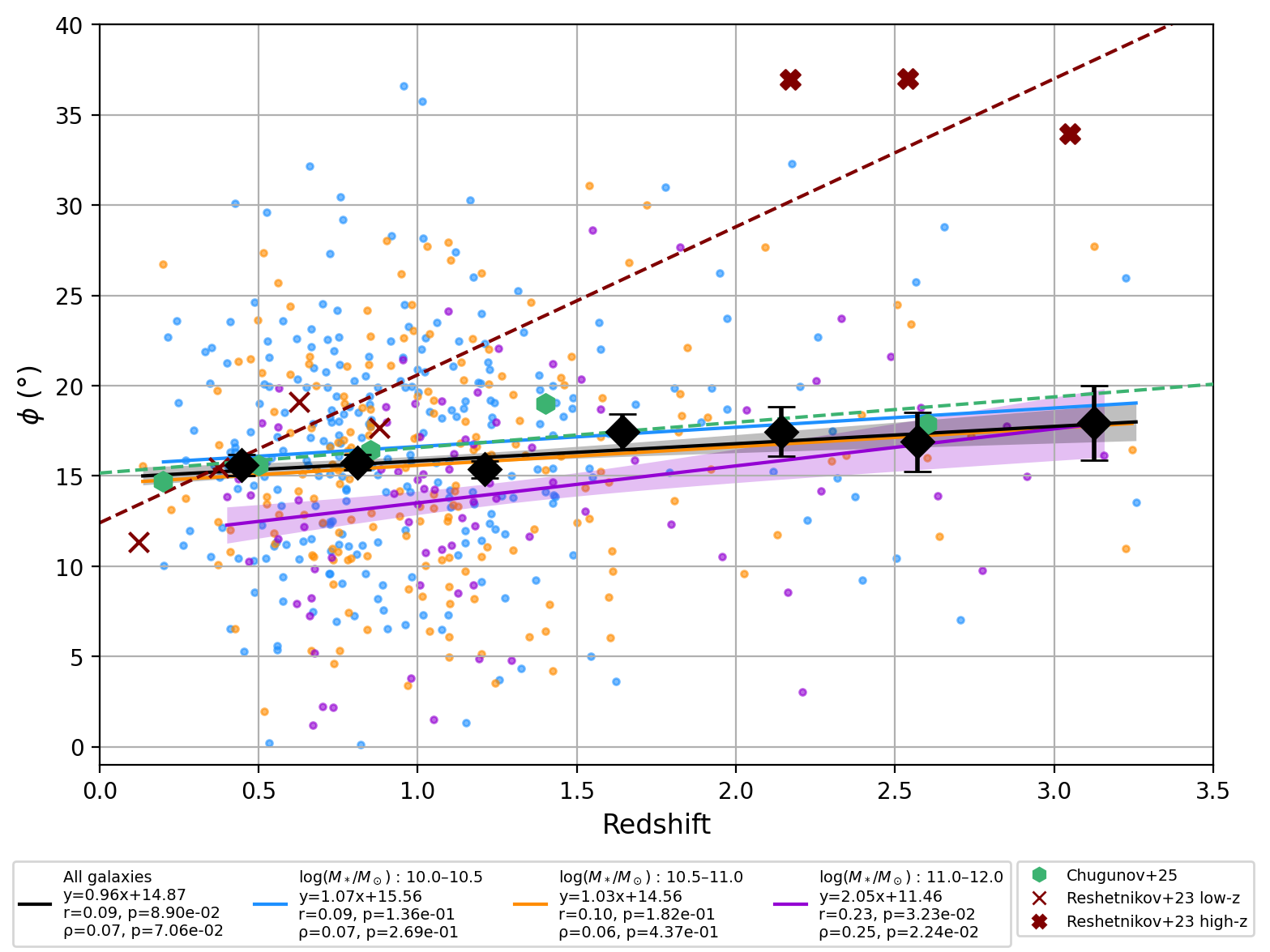}
\caption{Pitch angle as a function of redshift. The black diamonds and line represent the mean and standard deviation of the total sample. Values are colored by their stellar mass: log(M$_*$/M$_\odot$): 10-10.5 (blue), log(M$_*$/M$_\odot$): 10.5-11.0 (orange), log(M$_*$/M$_\odot$): 11.0-12.0 (purple). The shaded regions show a $1\sigma$ error on the mean. The solid lines show the best fit line for each mass bin. The best-fit line, Pearson and Spearman coefficients, with their p-values, are shown as well. Green hexagons and dashed line represent the mean from \cite{chugunov2025}. Thin maroon crosses and dashed line represent the mean from \cite{reshetnikov2023} and thick maroon crosses are 3 of the 4 high-redshift galaxies they analyzed.
\label{fig:redshift}}
\end{figure*}

In Figure \ref{fig:redshift}, we show the pitch angle vs. redshift relation. Our data shows that, overall (black diamonds and dashed line), there is a nearly flat trend in pitch angle across redshift. Both the Pearson and Spearman correlation coefficients are small, and the p-values are $>0.05$, indicating no correlation between pitch angle and redshift. We also tested this result by only using galaxies with spectroscopic redshifts and find no statistical difference.
We separated the galaxies into three mass bins: log(M$_*$/M$_\odot$)=[10.0-10.5, 10.5-11.0, 11.0-12.0]. The two smaller mass bins show a very similar relation to the overall trend (with 495 galaxies falling within these masses). The largest mass bin (98 galaxies) shows the largest increase with Pearson and Spearman correlation coefficients of 0.23 ($p=0.0323$) and 0.25 ($p=0.0224$), respectively, showing a weak correlation. This increase implies that the spiral arms of massive galaxies are winding up. We show a comparison with the high-redshift sample (green hexagons and dashed line) from \cite{chugunov2025} and the sample (thin maroon crosses) and individual measurements of high-redshift spiral galaxies (thick maroon crosses) from \cite{reshetnikov2023}.

\subsection{Pringle-Dobbs test}
\begin{figure}[ht!]
\includegraphics[width=1\linewidth]{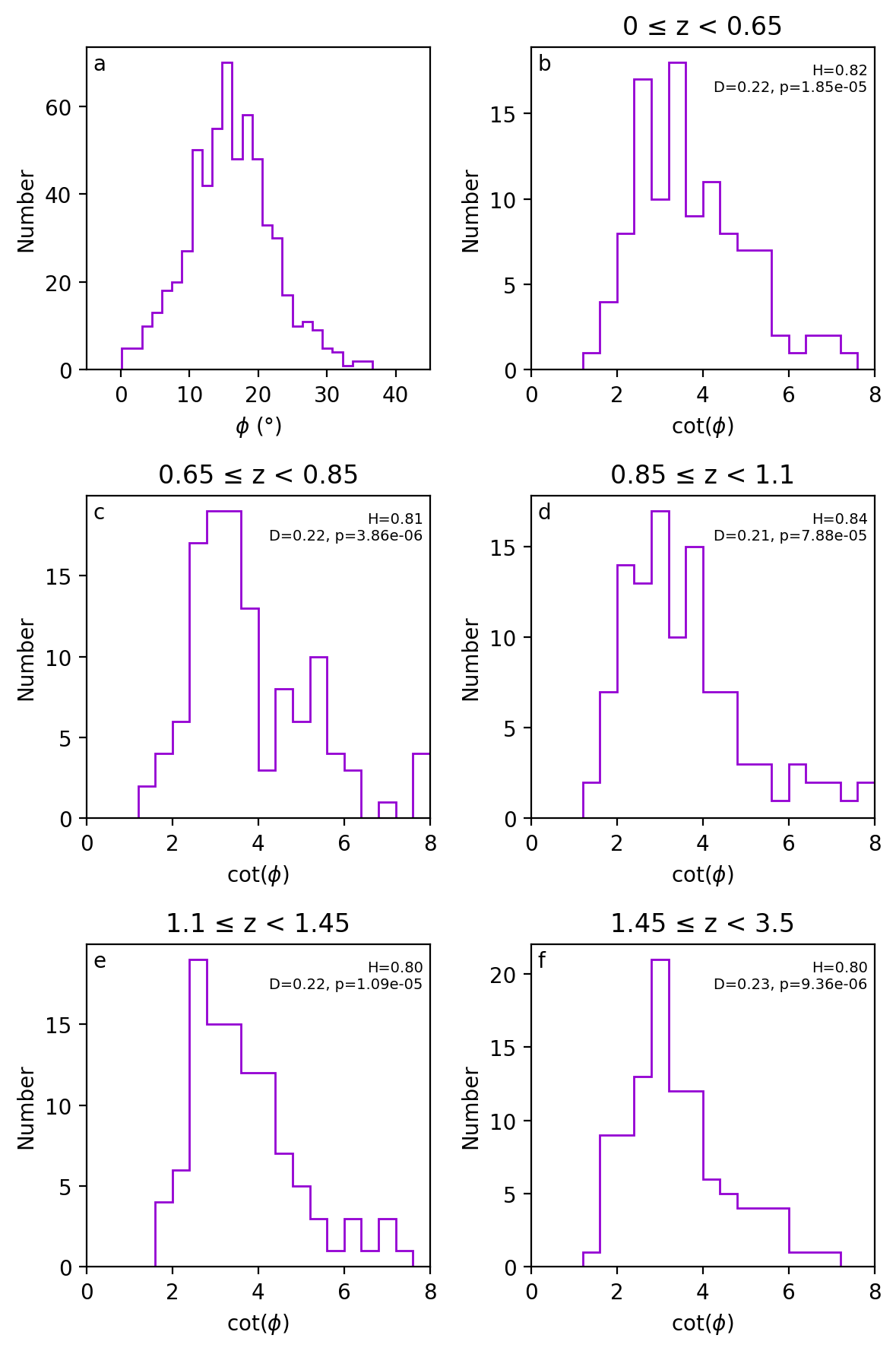}
\caption{(a) shows the distribution of pitch angles across our entire sample. (b)-(f) shows the Pringle-Dobbs test (distribution of $\cot\phi$).
\label{fig:pringle}}
\end{figure}

We show the distribution of pitch angles in the top left of Figure \ref{fig:pringle}. The distribution shows a strong peak around $\sim16^\circ$. Figure \ref{fig:pringle}b-f shows the distribution of the Pringle-Dobbs test \citep{pringle2019}. The Pringle-Dobbs test was proposed as a method to investigate the various theories of spiral structure formation. It states that if spiral arms are transient or tidal in nature, their pitch angle should decrease with time, where $\cot(\phi)\propto t$. Therefore, the distribution of the $\cot\phi$ should be uniform if arms are transient or tidal. This is contrary to the density wave theory, which predicts that pitch angles remain constant.

We binned the data so that there are approximately an equal number of galaxies per bin. In each bin, there is a large spike around $\cot\phi\sim3$, with no obvious uniform distribution in any bin. This trend is similar to that found by \cite{reshetnikov2022} at $z=0.55-1.1$ (see Figure 7 bottom row), which they attribute to different spiral arm mechanisms. However, the normalized Shannon entropy value (H) \citep{shannon1948} shows a fairly uniform distribution, where each bin has a value $\geq0.8$. There is a decrease in the entropy value as the redshift increases. When looking at the Kolmogorov-Smirnov (K-S) test, the $D$ values are on the edge of non-uniformity but have very small p-values ($\ll0.05)$ due to the sample size in each bin. Therefore, we find no uniformity in the $\cot\phi$ at any redshift. This points towards spiral arms being long-lived density waves, though other formation mechanisms are possible.

\subsection{Correlations between pitch angle and internal properties}
\begin{figure*}[ht!]
\includegraphics[width=1\linewidth]{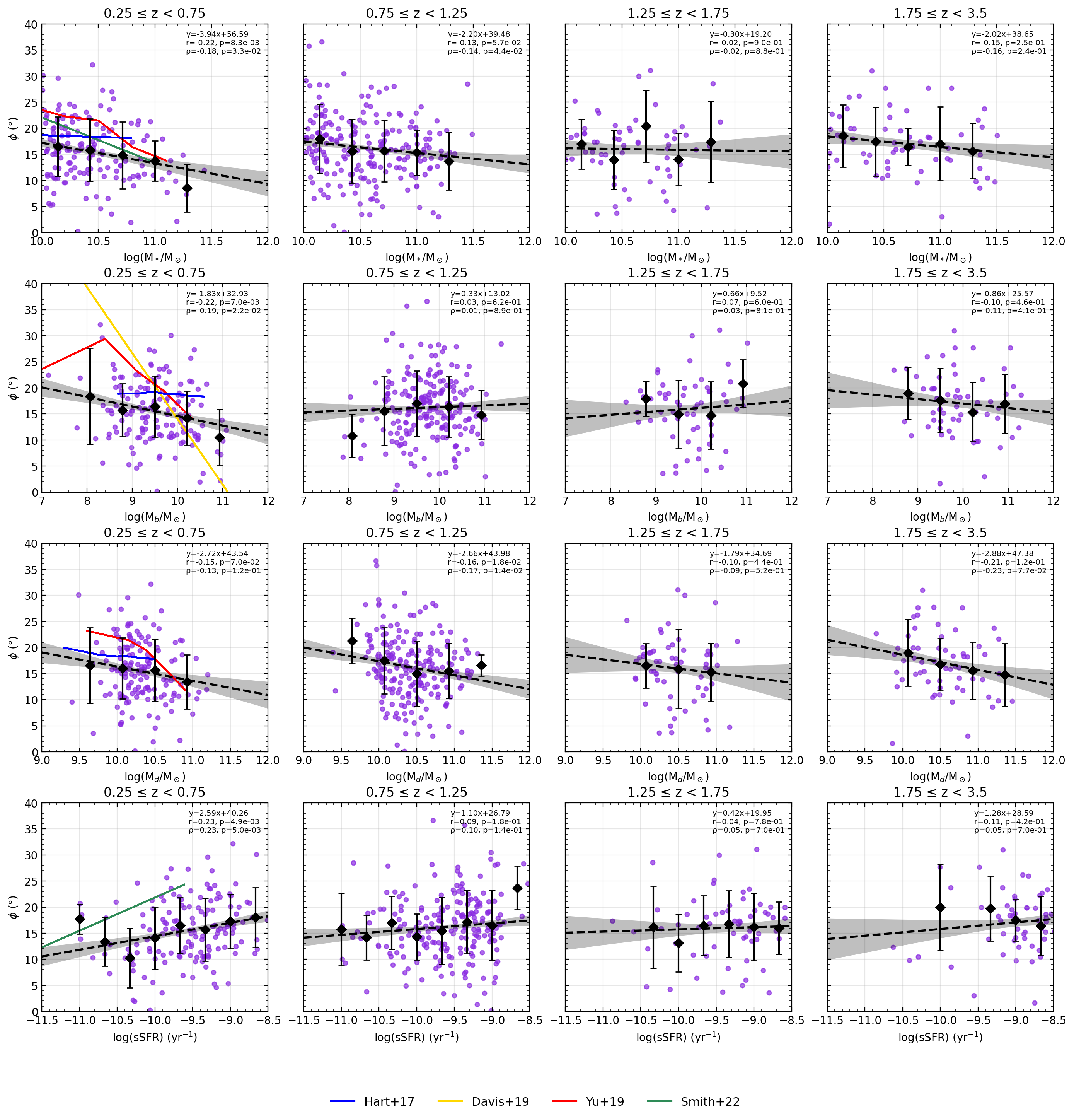}
\caption{First row: pitch angle as a function of total stellar mass. Black diamonds and dashed lines show mean and standard deviations. The shaded area shows the $1\sigma$ error on the mean. Best-fit line, Pearson, and Spearman correlation values, along with their p-values, are shown in the top right. Second row: pitch angle as a function of bulge mass. Third row: pitch angle as a function of disk mass. Fourth row: pitch angle as a function of sSFR. Results from low-redshift studies are shown in red \citep{yu2019}, blue \citep{hart2017}, yellow \citep{davis2019}, and green \citep{smith2022}.
\label{fig:mass}}
\end{figure*}

Figure \ref{fig:mass} shows the dependence of pitch angle on stellar mass (first row), bulge mass (second row), disk mass (third row), and sSFR (fourth row) in increasing redshift bins ($0.25\leq z<0.75$, $0.75\leq z<1.25$, $1.25\leq z<1.75$, $1.75\leq z<3.5$). For the stellar mass dependence, we see a weak but statistically significant negative trend in the first two bins ($r=-0.22$, $p=0.00083$ for $0.25\leq z<0.75$ and $\rho=-0.14$, $p=0.0044$ for $0.75\leq z<1.25$) showing that the more massive galaxies have smaller pitch angles. The last two redshift bins ($1.25\leq z<1.75$ and $1.75\leq z<3.5$) show a lot of scatter and no correlation. We show comparisons with \cite{hart2017} (blue line), \cite{yu2019} (red line), and \cite{smith2022} (green line).

The second row of Figure \ref{fig:mass} shows pitch angle versus bulge mass and we only observe a weak, but statistically significant, negative correlation in the $0.25\leq z<0.75$ bin ($r=-0.22$, $p=0.00073$) and no correlations in the other bins. Again, we show comparisons with \cite{hart2017} and \cite{yu2019}, but also \cite{davis2019} (yellow line). Likewise, the third row shows the pitch angle versus disk mass, and we again see a weak negative correlation in the second bin ($\rho=-0.16$, $p=0.0018$) and no correlation in the other three bins.

Lastly, we explore the correlation between pitch angle and sSFR in the bottom row of Figure \ref{fig:mass}. There is a weak correlation with sSFR in the first bin ($r=0.23$, $p=0.0049$), where spiral galaxies with looser arms have higher sSFR. Again, the other bins show no correlation. 



\subsection{Pitch angle vs. environment}
\begin{figure*}[ht!]
\includegraphics[width=1\linewidth]{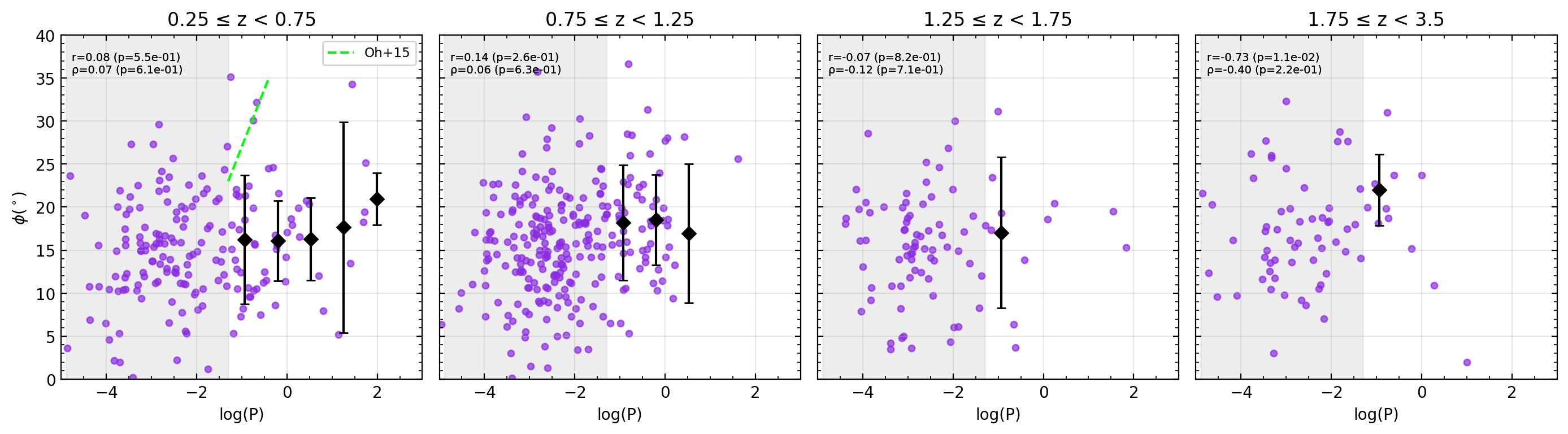}
\caption{Pitch angle as a function of the log of the tidal strength (P) exerted by a neighboring galaxy. We compared our results with \cite{oh2015} (lime dashed line). Our P values are much lower than those in \cite{oh2015} and we have shaded the region that falls outside their values.
\label{fig:environment}}
\end{figure*}

To determine if the pitch angle is dependent on the environment, we calculate the tidal strength of a companion galaxy as given by \cite{oh2015} and shown here:

\begin{equation}
    P=\left(\frac{M_{ptb}}{M_g}\right)\left(\frac{R_g}{r_{ptb}+R_{peri}}\right)^3
\end{equation}

where $M_{ptb}$ and $r_{ptb}$ are the neighbor galaxy's mass and size, respectively, $M_g$ and $R_g$ is the spiral galaxy's mass and size, respectively, and $R_{peri}$ is the distance to neighbor galaxy. We excluded $r_{ptb}$ as this value was much smaller compared to $R_{peri}$. $R_{peri}$ was calculated by finding all neighboring galaxies for each spiral galaxy within a radius of 1$\arcmin$ and a redshift range of $\Delta z<0.15$. For $R_g$, we calculated $7R_e/1.68$, to better match the values from \cite{oh2015}. We calculated the P value from all neighbors and used the largest value. We show the dependence of pitch angle on log(P) (Figure \ref{fig:environment}) and find no correlation at $z<1.25$. Our P values are lower than those reported in \cite{oh2015} (lime dashed line) and as such we have shaded everything below $P<0.05$ and only use the values above this threshold. In the two right columns, we do not have enough data points to draw a meaningful conclusion.


\section{Discussion} \label{sec:discussion}
As \cite{chugunov2025} pointed out, there are very few studies dissecting the nature of spiral galaxies at high-redshifts. Only recently have we begun to see an abundance of spirals at $z>1$ (e.g., \cite{kuhn2024,espejo_salcedo2025}). While we can make comparisons with other high-redshift studies, we are interested in how spiral galaxies change with redshift and how they compare with nearby spiral galaxies.

\subsection{Differences in pitch angle values}
Pitch angle values for nearby galaxies differ by several degrees. \cite{yu2019} found an average pitch angle of $\sim23^\circ$ for galaxies with $\sigma_c\lesssim100$ km s$^{-1}$ and later found a bimodality in the distribution with peaks at $\sim12^\circ$ and $\sim23^\circ$ \citep{yu2020}. They suggest that galaxies with lower pitch angles were produced by density waves and the galaxies with higher pitch angles are swing-amplified arms. Swing amplification can predict the pitch angle and arm number of a galaxy \citep{michikoshi2014}. A spiral arm, as a density wavelet, grows in amplitude until it reaches a maximum at a specific time, where the pitch angle can be calculated from the galactic shear. Swing amplification gives the unstable wavelength for density perturbations where the wavelength determines the dominant wave mode, corresponding to the number of arms formed \citep{michikoshi2014}. \cite{hart2017} found a mean pitch angle of $18^\circ$ and \cite{diaz_garcia2019} found a mean pitch angle of $\sim19^\circ$, whereas we found an average pitch angle of $z\sim16^\circ$.

The differences between pitch angle measurements at low redshifts and ours is most likely due to mass ranges and various measurement methods. Many studies use 2D Fourier analysis to measure pitch angles, though it only provides a single measurement for entire galaxies, rather than a value per arm (e.g., \cite{davis2012}). This is especially critical as it assumes that arms are symmetrical (higher redshift spiral galaxies are not as symmetrical as spiral galaxies at lower redshift \cite{chugunov2025}) and also doesn't account for variations along the arm, which can vary by $>10^\circ$ \citep{diaz_garcia2019,lingard2021}. \cite{chugunov2025} used a technique based on photometric decomposition and fitting a complex shape function to each arm. \cite{diaz_garcia2019}, \cite{yu2019}, and \cite{yu2020} used Fourier analysis, \cite{hart2017} used SpArcFiRe mixed with Galaxy Zoo, and \cite{reshetnikov2023} used a slicing method, where only part of the arm was measured. Our lower values, similar to \cite{hart2017}, could be due to using SpArcFiRe which calculates a length-weighted average (whereas Fourier-methods use flux-weighted), or it could be from increasing/decreasing SpArcFiRe parameters (i.e., contrast, clustering threshold). The galaxies from \cite{yu2019} and \cite{yu2020} have masses down to log($M_*/M_\odot)\approx8.5$ and with no cut to the inclination angle. Both \cite{hart2017} and \cite{diaz_garcia2019} use more face-on galaxies ($\alpha<65^\circ$) and masses as low as $\log(M_*/M_\odot)\sim9.45$ and $8.5$, respectively. We only focused on high-mass galaxies ($\log(M*/M_sol)\geq10$), which typically are more bulge-dominated and have tighter arms, rather than low-mass galaxies, which generally have smaller or no bulges and tend to have looser arms. The differences in pitch angles could also be attributed to the sample of spiral galaxies selected where flocculent galaxies typically have larger pitch angles than grand design or multi-armed. Galaxies with many arms in simulations exhibit higher pitch angles than 2-armed galaxies \citep{donghia2013,grand2013}. 

\subsection{Pitch angle evolution}
We observed that there is minimal change in the pitch angle overall as a function of redshift. This is contrary to what \cite{reshetnikov2023} and \cite{chugunov2025} found. \cite{reshetnikov2023} found an increase of $8.2^\circ$ per redshift (dashed maroon line) when only looking at massive (log($M_*/M_\odot)\geq10.5$) spiral galaxies (thin, maroon crosses). \cite{reshetnikov2023} also measured the pitch angles of four high-redshift galaxies (three of which we show in Figure \ref{fig:redshift}) and found them to have large pitch angles (thick, maroon crosses). Meanwhile, \cite{chugunov2025} found a much shallower slope, $0.5^\circ$ per Gigayear. Their data below $z<1.02$ comes from the COSMOS survey in the F814W filter (126 galaxies) and above this redshift, 33 galaxies from both the JADES and CEERS surveys ranging from 7-11 filters. Our trend (black line) is even shallower, though statistically insignificant. This suggests that spiral arms remain static over time, in line with the density wave theory. However, we did observe that the pitch angle of massive galaxies (log($M*/M_\odot)>11.0$) decreases towards lower redshift, similar to \cite{chugunov2025}, indicating that they wind up with time.




\subsection{Non-uniformity in the distribution of $\cot\phi$}
We found no uniformity in the distribution of $\cot\phi$ across any redshift bin when performing the Pringle-Dobbs test. The lack of uniformity implies that spiral arms form from density waves as the main mechanism, though some arms can still be transient. When we only consider our high mass galaxies, we still found non-uniformity across all redshift bins. We therefore found that this test did not provide conclusive results.

\cite{chugunov2025} performed this test on their entire sample, breaking their sample into roughly the same amount of galaxies per bin. They observed that their pitch angles were most uniform at low-redshift ($z\lesssim0.25$) and uniform at high-redshift ($z\gtrsim1.2$) based on their statistics. They speculated that mechanisms that produce spiral arms change over time: arms start off as tidal, then disappear while density waves remain; then when density waves fade, transient structures take over as the dominant factor. \cite{reshetnikov2023} also found a fairly uniform distribution towards low-redshift ($z\leq0.22$) and non-uniform distribution at $z=0.52-1.00$, also suggesting that there may be a change in mechanism.

\subsection{Internal Properties}
We found that there is a negative correlation between pitch angle and $M_*$ in only the two lowest redshift bins ($0.25\leq z < 0.75$ and $0.75\leq z < 1.25$). Both \cite{yu2019} (red line) and \cite{smith2022} (green line) found similar, though slightly steeper, trends to ours. This difference could be due to the samples themselves; \cite{yu2019} only had 79 relatively face-on galaxies while \cite{smith2022} had 2438 galaxies, both at $z<0.03$ and used ground-based observations. They both also used 2DFFT to measure their pitch angles (\cite{smith2022} adopted their pitch angle measurements from \cite{yu2020} for their analysis), whereas we used SpArcFiRe. The difference in slope could also be due to redshift evolution; our data shows that the slope increases with decreasing redshift (see top right panel of Figure \ref{fig:mass}). This negative correlation is consistent with the density wave theory but also can be explained by swing amplification (massive galaxies typically have higher shear rates which forces spiral arms to wind up tightly). Lastly, \cite{hart2017} (blue line) found no correlation between pitch angle and $M_*$, even though we used the same measurement program. The differences between the results could be due to their low redshift range ($0.02<z\leq0.055$), magnitude limit ($m_r\leq17.0$), mass range ($9.45<log(M_*/M_\odot\leq11.05$), and/or exclusion of strongly barred galaxies (unbarred and weakly barred only were included).

The negative correlation between pitch angle and bulge mass also supports the density wave theory; massive galaxies have a higher central mass concentration which leads to more tightly wound arms \citep{lin_shu1964,roberts1975}. \cite{davis2018} shows that the correlation between pitch angle and stellar mass is weaker than that between pitch angle and bulge mass suggesting that the mass of the bulge is the greater factor in the formation of spiral arms, consistent with the density wave theory. Furthermore, the "fundamental plane" of spiral structure shows a strong correlation where the tangent of the pitch angle is determined by the bulge mass and surface density of atomic hydrogen in the disk \citep{davis2015}. The bulge mass acts as a tension that forces the arms to become tighter as the central mass increases. We found a negative correlation at $0.25\leq z<0.75$, which is qualitatively consistent with \cite{davis2018}. However, \cite{davis2019} (yellow line) and \cite{yu2019} have much steeper slopes (at $z\sim0$). This correlation, again, could have been established at an earlier time and become steeper with cosmic time. 

We observed similar correlations between the pitch angle and disk mass as we did with total stellar mass, though only statistically significant at $0.75\leq z<1.25$. Our trend follows more closely that of \cite{hart2017} whereas \cite{yu2019} has a steeper slope.




We found a positive correlation between pitch angle and sSFR only at $0.25\leq z<0.75$. \cite{smith2022} also observed a similar, though stronger, relation. This correlation, though weak at low-redshift, further supports the fundamental plane where a higher gas density (and higher sSFR) loosens spiral arms \citep{davis2015}.


\subsection{External Factors}
We found that the pitch angle is not dependent on tidal strength. Our values for the tidal strength are much lower than the values found by \cite{oh2015}. Our slope is relatively flat compared to theirs therefore we can infer that the environment, by way of the tidal strength, does not influence the pitch angle. \cite{smith2022} found that spiral galaxies in clusters have smaller pitch angles, though this is attributed to the larger central concentration and when that is taken into account, there is no correlation between pitch angle and environment. This implies that the tightness of arms is dependent on internal, rather than external factors.


\subsection{Spiral arms as density waves}
We found no correlations with any properties at $z\geq1.25$. At this redshift, disk galaxies are most likely dynamically hot (e.g., \cite{genzel2020,nelson2024}) and clumpy (e.g., \cite{guo2012,guo2015,guo2018,kalita2024}). In both observations and simulations, merger rates increase with redshift (e.g., \cite{cibinel2019,pillepich2019}). Therefore, it is likely that tidal arms formed through these interactions and/or since these galaxies were hot, clumpy, and turbulent, spiral arms could form via swing amplification and be transient in nature.

By $z\sim0.75$, disks have become settled, cold, thin, less clumpy, and more organized (e.g., \cite{kassin2012,genzel2017,tiley2021,martin2023,kuhn2024,espejo_salcedo2025}). Here, we start to see galaxies build up their mass and long-lived density waves take over. We see this through the non-uniform distribution of $\cot\phi$ in Fig. \ref{fig:pringle} and Fig. \ref{fig:mass}, but also the constant pitch angle over time (Fig. \ref{fig:redshift}). Interactions can still occur and loosen spiral arms, though it is not the primary factor. 

Below $z<0.75$, the pitch angle--tidal strength relation grows weaker while bulges have become prominent and the pitch angle--bulge mass relation has become stronger. At $z\sim0$, this relation is even stronger as evidenced by \cite{davis2019} and \cite{yu2019} (Fig. \ref{fig:mass} first column second row). Spiral arms are well described by the density wave theory. The most massive galaxies are the first to settle down to a stable disk and can slowly wind up their arms (Fig. \ref{fig:redshift}). This could be due to non-steady global stellar waves, where the pattern speed of the waves changes with the radius \citep{elmegreen2011}.

\begin{figure}[h]
\includegraphics[width=1\linewidth]{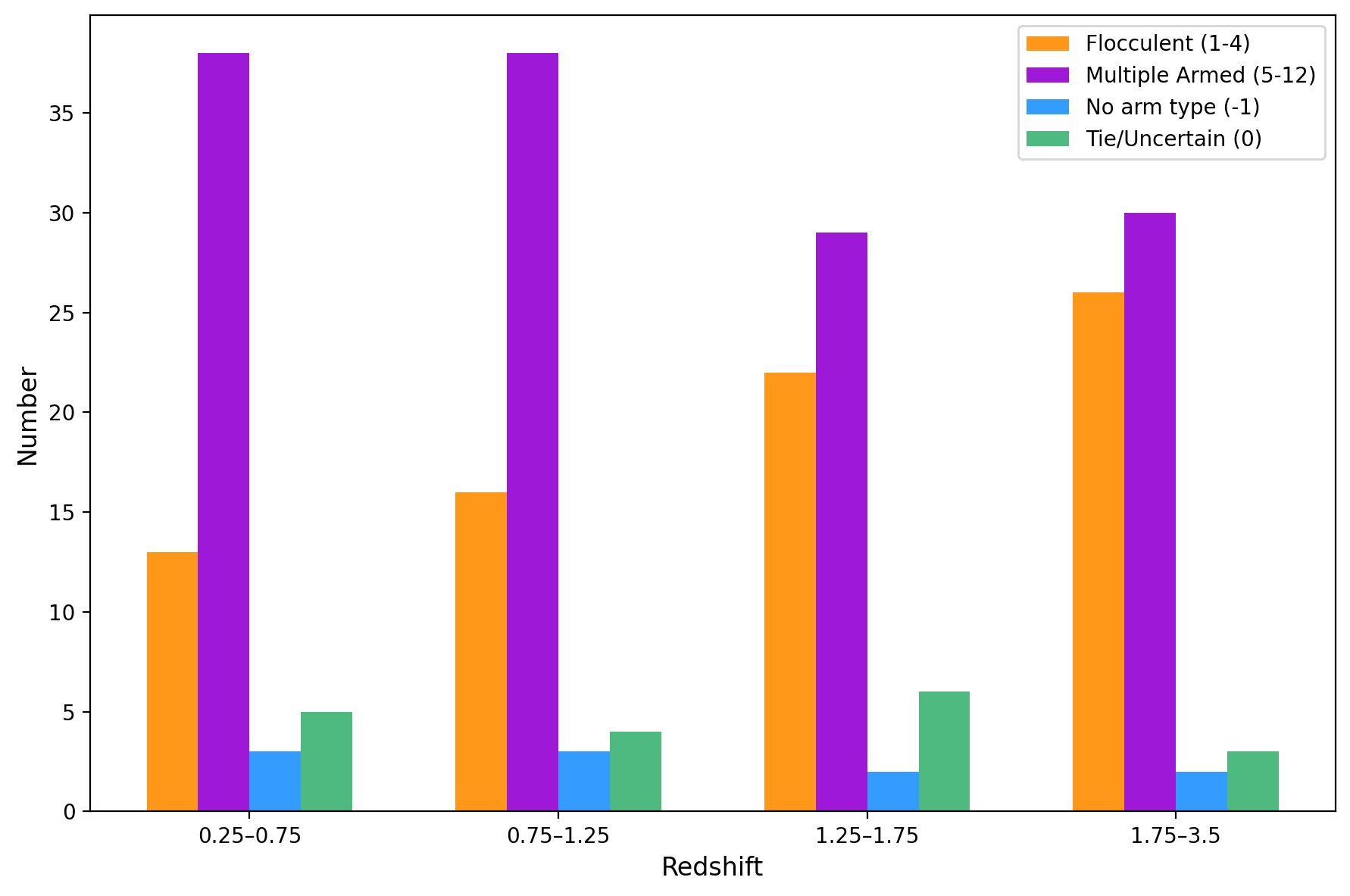}
\caption{Results of our arm classification. The number of flocculent galaxies (orange) decreases with cosmic time whereas multiple-armed galaxies (purple) slightly increase. Blue bars represent the category where the classifiers could not see a clear arm class and green bars represent the galaxies where there was no arm class majority.}
\label{fig:arms}
\end{figure}

As further support for a change at $z\sim1.25$, we classified a subset of galaxies based on arm morphology. We selected a sample of 60 galaxies in four redshift bins to determine arm type. We based our arm classes on the 12 arm classification system from \cite{elmegreen1982}. We had 5 classifiers visually inspect each of the randomly drawn galaxies. The classifiers were aided with both the written description and visual examples for each class. Typically, arms are categorized into 2 or 3 categories: flocculent, multi-arm, and grand design. In this paper, we only use 2 categories (1-4: flocculent and 5-12: multiple armed). The results were broken down into 4 categories where they were grouped into a singular category if they had 3+ votes: arm class could not be identified (-1), flocculent arms (1), multiple armed (2), and if there was a tie or no majority (0). Results are shown in Figure \ref{fig:arms}.


We found that multiple-armed galaxies remain fairly steady at $1.25\leq z<3.5$ and increase slightly at $0.25\leq z<1.25$. The number of flocculent galaxies increases with increasing redshift. Very rarely were we not able to classify arms. We observed that the number of galaxies in all four types is nearly the same for the first two bins. Observationally, there is a difference between the second and third redshift bins, though it is not of statistical significance. The results from our contingency test do not show a statistical significance between the first and fourth bins ($\chi^2=3.32$, $p=0.0684$) for multiple armed and flocculent galaxies. This decrease in flocculent type galaxies shows that spiral galaxies settle over time and become more organized. As they settle and become cold, their bulges grow and they are able to maintain their spiral arms for longer periods of time, consistent with density waves.

\section{Summary} \label{sec:summary}
In this paper, we identified spiral galaxies from CEERS and JADES using the machine learning algorithm Zoobot. We then measured their pitch angles using SpArcFiRe. Here we summarize our results:
\begin{enumerate}
    \item We identified 657 spiral galaxies at $0.0<z<3.5$, with 326 being at $z>1$. We obtained pitch angle measurements for 593 spiral galaxies. This is the largest study of high-redshift spiral galaxies to date.
        
    \item The overall distribution of pitch angles shows an average around $16^\circ$ with a few above 30$^\circ$. When applying the Pringle-Dobbs test, we found no uniformity at any redshift bin. Therefore, this suggests that spiral arms cannot purely form from transient or tidal interactions.
    
    \item We found that the overall pitch angle remains mostly constant with time. The highest mass galaxies (log$(M_*/M_\odot)=11-12$) show a change in pitch angle indicating that their spiral arms become tighter with cosmic time. This is in agreement with previous results from literature.
    
    \item We observed a weak, but statistically significant, negative correlation between the pitch angle and total stellar mass at $z<1.25$ and between pitch angle and disk mass at $0.74\leq z<1.25$. We also observed a negative correlation between the pitch angle and bulge mass, but only in the lowest redshift bin ($0.25\leq z<0.75$). Our result agrees with several low-redshift studies using various methods to measure the pitch angle. However, the slopes of our correlations are not as steep as those at $z\sim0$ in the literature, possibly indicating a redshift evolution. Pitch angle is also positively, though weakly, correlated with sSFR at $0.25\leq z<0.75$.
    
    \item We measured the tidal strength applied by companion galaxies to our sample. We found that the tidal strength effect in our sample is significantly smaller than those discussed in the literature and the pitch angle has no correlation with the tidal strength. Therefore, external factors are not as important as internal factors regarding the tightness of spiral arms.

    \item The number of flocculent galaxies increases with increasing redshift while multiple armed galaxies decrease with increasing redshift. This indicates that disk galaxies become more stable over time and are able to maintain spiral arms. 

    

\end{enumerate}

Overall, we infer the following spiral arm formation mechanisms from our results. The lack of environmental dependence implies that internal disk properties dominate across all redshifts. There is a transition epoch at $z\sim1$.
At $z\gtrsim1$, spirals are not governed by global density wave modes; while 
at $z\lesssim1$, spirals are regulated by global gravitational potential. A more detailed follow up study of the internal structures would shed more light on the mechanisms responsible for creating and maintaining arm structures.

\begin{acknowledgments}
We thank the anonymous referee for their valuable comments that lead to the improvement of this paper. This work is supported by NASA-Missouri Space Grant Consortium  No. 80NSSC20M0100 and the
University of Missouri Research Council grant URC-23-036. All the {\it JWST} data used in this paper can be found in MAST: \dataset[10.17909/xm2a-3731]{https://doi.org/10.17909/xm2a-3731}.

\end{acknowledgments}

\bibliography{sample7}{}
\bibliographystyle{aasjournalv7}



\end{document}